\documentclass[12pt]{iopart}
\usepackage{iopams}
\usepackage{graphicx}
\usepackage{psfrag}
\usepackage{dcolumn}
\usepackage{latexsym}
\usepackage{amssymb,latexsym,mathrsfs}
\usepackage{amsfonts}

\newcommand{\beq}{\begin{equation}}
\newcommand{\eeq}{\end{equation}}
\newcommand{\beqn}{\begin{eqnarray}}
\newcommand{\eeqn}{\end{eqnarray}}
\newcommand{\bearr}{\begin{array}}
\newcommand{\enarr}{\end{array}}

\newcommand{\ra}{\rangle}
\newcommand{\la}{\langle}

\def\bea{\begin{eqnarray}}
\def\eea{\end{eqnarray}}
\def\ba{\begin{array}}
\def\ea{\end{array}}

\begin{document}

\title{ Asymmetric Simple Exclusion Process on a Cayley Tree }
\author{ Mahashweta Basu and P. K. Mohanty}

 \address{Theoretical Condensed Matter Physics Division,\\ Saha Institute of Nuclear Physics,
1/AF Bidhan Nagar, Kolkata, 700064 India.}
\ead{mahashweta.basu@saha.ac.in and pk.mohanty@saha.ac.in}



\begin{abstract}
  We study  the asymmetric exclusion process  on a regular Cayley tree with arbitrary co-ordination 
number.  In this model particles can enter the system  only at the parent site and exit from one 
of the sites at  the last level.  In the bulk they move from the occupied sites to one of their
unoccupied downward neighbours, chosen randomly. We show that the steady state current that  flow 
from one level to the next is independent of the exit rate,  and  increase monotonically with the 
entry rate and the co-ordination number. Unlike TASEP, the model  has only one  phase and  
the density profile show no  boundary layers. We argue that in blood, air or water circulations 
systems branching is essential to  maintain a free flow  within the system which is independent 
of  exit rates.
\end{abstract}
\maketitle

 Exclusion processes \cite{exclusion} has been studied extensively as  the paradigm models 
of non-equilibrium phase transitions.  They exhibit reach variety of phases, 
phase-coexistence,  shock-profiles and non-trivial  boundary layers.  Some variations of 
these models are  exactly solvable \cite{exact} on a one  dimensional lattice, which 
provide deep understanding  of  non-equilibrium transport, traffic and jamming.  
However, very little is known  about the systems beyond one spatial dimension. In particular, 
transport in  irregular structures, like networks has  been a recent 
topic of interest \cite{transnet}.  In a generic undirected network particle can enter 
or exit at  any arbitrary sites. Again, the presence of loops in the generic 
networks, also make  the study of  particle transport difficult. A prototype 
network is a Cayley tree, where the direction of transport, the entry and  exit
points are  well defined. Absence  of loops make the study relatively  simpler. 
Again,  several physical systems like, water transport in trees,  transport of 
nutrients in  blood-circulation system \cite{blood}, transport of antibody in idiotypic 
networks in immune system\cite{immune},   air circulation in lung \cite{lung}, and 
flow on disordered networks\cite{Stinchcombe} are strikingly similar to this model 
system of Cayley trees.

\begin{figure}[h]
 \centering
\includegraphics[width=5.5cm]{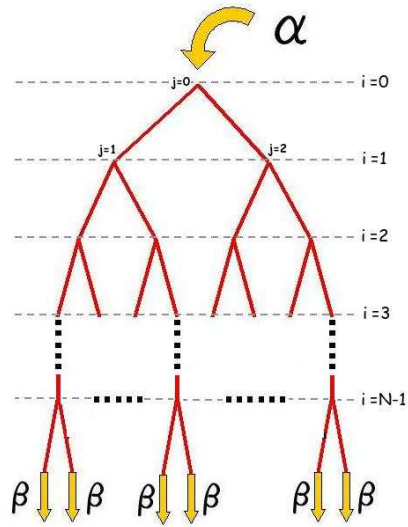}
 \caption{Rooted Cayley tree  with $q=2$. There are $N$ levels and $M=(2^N -1)$ sites, 
labeled  by $i=0,1\dots,N-1$ and $j=0,1\dots,M$ respectively. Particle enter at the parent 
site $j=0$ with rate $\alpha$ and exit from any of the sites at the last level $N-1$ with rate $\beta$.}
 \label{fig:model}
\end{figure}

  In  this Letter we study transport in a Cayley tree with arbitrary co-ordination
number. Particles can enter the Cayley tree only at the parent site  with 
rate $\alpha$ and  are allowed to leave from any of the sites at the last level 
with rate  $\beta$. In the bulk, when allowed by hard-core interaction,
the particle can  move  to one of its downward neighbour chosen randomly.  
The resulting current that flow between  neighboring levels is found to be independent of 
the exit rate $\beta$ when $q\ge 2$. Further,  the current is found to be larger than that of the 
TASEP in one dimension for any value of $(\alpha, \beta)$. The average density at  each 
level, except the last one, do not   show any finite size correction and decay 
exponentially as one moves away from the parent site.  
In the last level, the density depends both on $\alpha$ and $\beta $ and 
decays exponentially  with the system size. We did a mean field analysis, which 
correctly reproduces all these features. 


First the model. Cayley tree, is a connected cycle-free graph. A $N$-level 
Cayley tree, labeled as $i=0,\dots N-1$, with co-ordination number $z= q+1$  can be 
generated  by starting form the parent level $i=0$ with one 
single site $j=0$ called root.  The next generation of sites  are then created 
iteratively, where  each site in level $i$  is connected to  $q$ 
new sites  in the next level $i+1$.  Thus level $i$ has $q^i$ sites and  
the resulting network has total $M =(q^N-1)/(q-1)$ sites, 
labeled by $j=0,\dots M$.

Note, that the  Cayley tree with $q=1$  is a one dimensional lattice with 
$N$ sites. The first non-trivial structure  can be obtained for $q=2$, which is 
described in the Fig. \ref{fig:model}.  Here, in every level, each site is 
connected to two downward neighbours in the next level. There are $M=(2^N -1)$ sites
which are labeled by $j$ increasing from top to bottom and left to right. 
Further, each site  of a Cayley tree can either be vacant or occupied by at most 
one particle. Correspondingly 
we define  a variable $s_j = 1, 0$ at site $j$,  which represents a presence or absence 
of a particle respectively.  These hardcore particles 
flow downwards (from level $i$ to  level $i+1$)  with the following rules. 
A particle present in a level $i$, at  the site say $j$, will move to 
its neighbour $j^\prime$ in the next lower level $i+1$ only when $s_{j^\prime}=0$. 
If more than one neighbour of $j$ is vacant, one of them is chosen randomly for 
particle transfer. In the following, the bulk dynamics for $q=2$ is described  schematically. 

\vspace{.2cm}
 \hspace*{2.9 cm} 
\setlength{\unitlength}{.4 cm}
\begin{picture}(1,1) 
  \put(0, 0)  {\line(-1,-1){1}}
 \put(0, 0){\line(1, -1){1}} 
\put(-0.2, .3){1}
\put(-1.6,-1.2){$1$}
\put(1.1,-1.2) {$0$}
\put(2.6,.1) {$1$}
\put(2.1,-.4){$\longrightarrow$}
\end{picture}
\hspace*{1.7 cm} 
\begin{picture}(1, 1)
  \put(0, 0)  {\line(-1,-1){1}}
  \put(0, 0){\line(1, -1){1}} 
\put(-0.2, 0.3){$0$}
\put(-1.6,-1.2){$1$}
\put(1.1,-1.2) {$1$}
\put(2.4,-.4) {$;$}
\end{picture}
\hspace*{1.5cm} 
\begin{picture}(1, 1)
  \put(0, 0)  {\line(-1,-1){1}}
  \put(0, 0){\line(1, -1){1}} 
\put(-0.2, 0.3){$1$}
\put(-1.6,-1.2){$0$}
\put(1.1,-1.2) {$1$}
\put(2.6,.1) {$1$}
\put(2.1,-.4) {$\longrightarrow$}
\end{picture}
\hspace*{1.7 cm} 
\begin{picture}(1, 1)
  \put(0, 0)  {\line(-1,-1){1}}
  \put(0, 0){\line(1, -1){1}} 
\put(-0.2, 0.3){$0$}
\put(-1.6,-1.2){$1$}
\put(1.1,-1.2) {$1$}
\end{picture}
 \beq \label{eq:dyn1}\eeq

 \hspace*{2.9 cm} 
\begin{picture}(1,1) 
  \put(0, 0)  {\line(-1,-1){1}}
 \put(0, 0){\line(1, -1){1}} 
\put(-0.2, .3){1}
\put(-1.6,-1.2){$0$}
\put(1.1,-1.2) {$0$}
\put(2.6,.4) {$\frac 1 2$}
\put(2.1,-.4){$\longrightarrow$}
\end{picture}
\hspace*{1.7 cm} 
\begin{picture}(1, 1)
  \put(0, 0)  {\line(-1,-1){1}}
  \put(0, 0){\line(1, -1){1}} 
\put(-0.2, 0.3){$0$}
\put(-1.6,-1.2){$1$}
\put(1.1,-1.2) {$0$}
\put(2.4,-.4) {$;$}
\end{picture}
\hspace*{1.5 cm} 
\begin{picture}(1, 1)
  \put(0, 0)  {\line(-1,-1){1}}
  \put(0, 0){\line(1, -1){1}} 
\put(-0.2, 0.3){$1$}
\put(-1.6,-1.2){$0$}
\put(1.1,-1.2) {$0$}
\put(2.6,.4) {$\frac 1 2$}
\put(2.1,-.4){$\longrightarrow$}
\end{picture}
\hspace*{1.7 cm} 
\begin{picture}(1, 1)
  \put(0, 0)  {\line(-1,-1){1}}
  \put(0, 0){\line(1, -1){1}} 
\put(-0.2, 0.3){$0$}
\put(-1.6,-1.2){$0$}
\put(1.1,-1.2) {$1$}
\end{picture}
 \beq \label{eq:dyn2}\eeq

Flow of particles 
is maintained  in the system by the in- and out-fluxes of particles  at boundaries. 
A particle can enter the system  with rate $\alpha$ if the parent site  $j=0$  is  
unoccupied ($s_j=0$).  
Any  particle from  the  $(q^N-1)$ sites at the  boundary level $N-1$ can leave the system 
with rate $\beta$. Note,  that for $q=1$, each site has only one downward neighbour and  
the dynamics is simply $10\to 01$. Such an exclusion process, namely Totally Asymmetric 
Simple Exclusion Process (TASEP)\cite{tasep} on a one dimensional lattice, has been studied
extensively  for its  application in  vehicular traffic,  cellular transport by motor-proteins,
etc.. TASEP has  been solved exactly\cite{derrida, mpa}, where it shows  novel properties  like 
boundary driven phase transition,  shock formation and propagation, condensation, and jamming. 

 It is important to note that the bulk dynamics in the Cayley tree is particle conserving.  
In steady state, the average current that flow between neighbouring 
levels $i$ and $i+1$ is expected to be constant, say  $I_q$.  Thus the average current 
on each link that connects a  site  at level  $i$ with  its neighbour at level $i+1$ 
is given by  
\beq J_{i \rightarrow i+1} \equiv J_i = \frac{I_q}{q^{i+1}}.\label{eq:Ji}\eeq
All configurations where a site in $i^{th}$ level is occupied 
and  at least one of  its $q$neighbour in   $(i+1)^{th}$ level is empty, contributes 
to the  flow of particles. Thus, fraction of such configurations in steady state 
is $q J_i$.  Here, the  factor $q$   takes care of the 
fact that the  average current $I_q$ is shared among $q$ bonds. 
For example, when $q=2$, we have\\
 \hspace*{4.1 cm}  
\setlength{\unitlength}{.4 cm}
\begin{picture}(1, 1) 
  \put(0, 0)  {\line(-1,-1){1}}
  \put(0, 0){\line(1, -1){1}} 
\put(0, 0){1}
\put(-2.,-1){$\la 1$}
\put(1,-1) {$0\ra$}
\put(-5.,-1){ $2 J_i =$}
\end{picture}
\hspace*{1.3 cm} 
\begin{picture}(1, 1)
  \put(0, 0)  {\line(-1,-1){1}}
  \put(0, 0){\line(1, -1){1}} 
\put(0, 0){1}
\put(-2.9,-1){$+$}
\put(-2.,-1){$\la 0$}
\put(1,-1) {$1\ra$}
\end{picture}
\hspace*{1.3 cm} 
\begin{picture}(1, 1)
  \put(0, 0)  {\line(-1,-1){1}}
  \put(0, 0){\line(1, -1){1}} 
\put(0, 0){1}
\put(-2.9,-1){$+$}
\put(-2.,-1){$\la0$}
\put(1,-1) {$0\ra,$}
\end{picture}
 \beq \label{eq:bondJ}\eeq
where $\la\dots\ra$ represents  the steady state averages. 
Similarly the total density at level $i$  
\beq 
\phi_i=\sum_{j \in i}\la s_i \ra. 
\label{eq:phi}
\eeq
is the steady state average of the occupied sites  belongs to level $i$.

In the following we use  the mean field theory (MFT) where both the fluctuations of particle density at  
the individual sites and  variation  of densities among  sites in the same level, are neglected. 
At level $i$, thus, every site is assumed to have an average density 
 \beq \rho_i={\phi_i\over q^i},\label{eq:rhoi} \eeq
where  $\phi_i$ is the total density at level $i$. 
Since a site  of level $i$ is occupied  with probability 
$\rho_i$ (and is vacant with probability $1-\rho_i$), the  average current through 
the bonds is given by,
\beq
q J_i = \rho_i \left(  1-\rho_{i+1} ^q \right),
\label{eq:J}
\eeq 
where$(1-\rho_{i+1} ^q)$ is  the meanfield probability that at least 
one site of $(i+1)^{th}$ level is unoccupied. The factor $q$ in front of $J_i$ 
ensures that  the outgoing current flows through $q$ bonds.

The conservation of particle density in the bulk of the system 
leads to a continuity equation 
\bea
{d\rho_i \over dt}&=& J_{(i-1)\to i}  -  qJ_{i\to (i+1)}= J_{i-1}-qJ_i\cr
&=& {1\over q}\rho_{i-1}(1-\rho_i^q)-\rho_i(1-\rho_{i+1}^q).
\label{eq:rhoN-1on}
\eea 
These equations must be supplemented by the following 
boundary conditions. First, at the root $j=0$, where particle enters to the system, 
we have 
\begin{equation}
 {d\rho_0 \over dt} = \alpha(1-\rho_0)- qJ_0 = \alpha(1-\rho_0)-\rho_0(1-\rho_1^q).
\label{eq:eq0}
\end{equation}
 Similarly at the last level $i=N-1$, 
\bea
 {d\rho_{N-1} \over dt} &=&   J_{N-2} - \beta\rho_{N-1}\cr  &=& {1\over q}\rho_{N-2}(1-\rho_{N-1}^q)-\beta\rho_{N-1}.
\label{eq:eqN}
\eea

From  Eq. \ref{eq:rhoN-1on} it is clear that, in the steady state  the bulk current is 
 $J_i=J_{i-1}/q$, which can be iterated  to give 
\beq J_i = {J_0\over q^i}.\eeq  A comparison of this equation with  Eq. (\ref{eq:Ji})
results, $I_q=qJ_0$. Thus  Eqs. (\ref{eq:J}) and (\ref{eq:Ji}) provides  an 
iterative equation for the density, 
\begin{equation}
 \rho_{i+1}=  \sqrt[q]{1-{I_q\over q^i\rho_i}}  
\label{eq:e7}
\end{equation}
 \begin{figure}
\centering
\includegraphics[height=6cm]{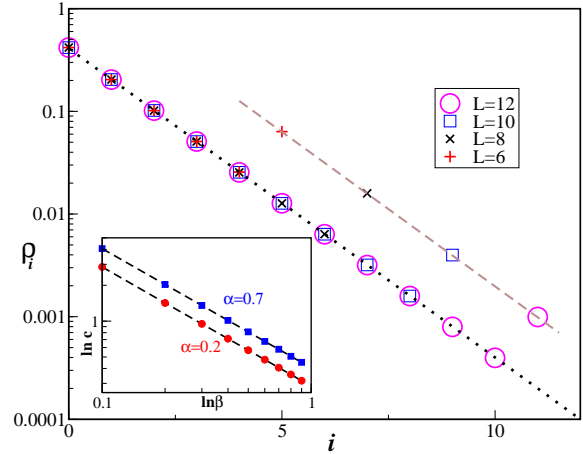}
\caption{ Density profile for $q=2$ obtained from Monte-Carlo simulations (symbols) 
with $\alpha=0.7$ and  $\beta= 0.2$ are  
compared with  Eq. \ref{eq:rhor0} (dotted  line) where $\rho_0=0.422$  was calculated using
\ref{eq:theta}.  Clearly there are no finite size corrections. 
The densities of the last level $\rho_{N-1}$  also follow Eq. \ref{eq:rhoN-1} (dashed line).
In  the inset we compare $c$ versus $\beta$  obtained from simulations (symbols) for 
$\alpha=0.4,0.7$ with Eq. \ref{eq:rhoN-1}.
}
 \label{fig:rhoN}
\end{figure}

First let us discuse the case $q=1$ ($i.e.$ TASEP) which is conceptually 
different from  other cases $q\ge 2$ as the right  hand side of the above map
for $q=1$ do not depend explicitly on $i$ ( $i.e.$  $\rho_{i+1}=1-I_1/\rho_i$).
In TASEP both, the density profile and the current  show macroscopic  
changes as one vary the boundary rates $\alpha$ and $\beta$, resulting  in 
three different phases : (i) the high density phase($\alpha>\beta,\beta<\frac 1 2$), 
(ii) the low density phase ($\alpha<\frac 1 2>, \beta>\alpha$), and (iii) the maximal current 
phase  ($\alpha>\frac 1 2, \beta>\frac 1 2$). TASEP is an exactly solvable 
model\cite{derrida}, however the meanfield  analysis of  Eq. \ref{eq:e7} with 
$q=1$  is known\cite{mpa} to produce correct  phase diagram and the  density 
profiles. 

   One must treat the $q\ge2$ case  separately as the  function in the right 
hand side of Eq. \ref{eq:e7} changes in each iteration. The existence of the 
fixed points in such maps are not quite obvious.  Let us take 
the large $i$ limit and  re-write the map in terms   of $\phi_i= q^i \rho_i$, 
which is an ${\cal O}(1)$ quantity. Thus,
\begin{equation}
 \phi_{i+1}=q^{i+1}\sqrt[q]{1-{I_q \over \phi_i}}.
\label{eq:e8}
\end{equation}
The fixed point of the map is then $\phi_i= \phi^*$, given by
\begin{equation}
 \phi^*= I_q.
\label{eq:fp}
\end{equation}
Thus in the bulk we have a  constant  level density $\phi_i=I_q$.
This is similar to the case $q=1(TASEP)$, where the bulk density  was 
found to be a constant. However,  the density profile in TASEP show a 
boundary layer at both ends. Since the fixed point solution $\phi_i=I_q$ 
is  expected to hold only for large $i$, it is not clear if the total 
density  $\phi_i$ has a boundary layer for $q\ge2$. The detailed mean 
field analysis (discussed below) show that the boundary  layers are in 
fact absent for $q\ge 2$. 

To calculate $I_q$ we  use the boundary 
condition (\ref{eq:eq0}),
\begin{equation}
I_q=\alpha(1-\rho_0).
\label{eq:Jr0}
\end{equation}
Finally,  Eqs.  (\ref{eq:rhoi}), (\ref{eq:fp}) and  (\ref{eq:Jr0}) can be used 
to  obtain the average level density, 
\beq
\rho_i = \frac{\phi_i}{q^i}= \frac{\alpha(1-\rho_0)}{q^i} = \frac{I_q}{q^i}  .
\label{eq:rhor0}
\eeq 
Thus, the mean field  densities $\rho_i$  and the current $I_q$  are expressed as   a single parameter functions of $\rho_0$ (which is same as $\phi_0$). To calculate  $\rho_0$  we use the  boundary condition (\ref{eq:eq0}). It demands that, in the steady state $\rho_1=  \sqrt[q]{1-I_q/\rho_0}$,  whereas from  
Eq. \ref{eq:rhor0} we have $\rho_1= I_q/q$. Hence,  
 \beq
 \sqrt[q]{1- {\alpha (1-\rho_0) \over \rho_0}}= {\alpha(1-\rho_0) \over q}.\label{eq:phi0}
\eeq
The above $(q+1)^{th}$ order equation in $\rho_0$ can not  have a close form 
solution for  $q> 3$. However, numerical solution for any given $\alpha$ and $q$ can 
be obtained  with high accuracy. Analytical solutions  of  Eq. \ref{eq:phi0} can 
be obtained for $q=2,3$. For example,  when  $q=2$   we have,  
\beq \rho_0={2\over3}[1-{\sqrt{\alpha^2+12\alpha+12}\over \alpha}cos({\pi+\theta\over3})], \nonumber \eeq
\beq{\rm where}~~tan\theta={6\sqrt{6}\over\alpha}{\sqrt{\alpha^4+18\alpha^3+20\alpha^2
+24\alpha +8}\over36-18\alpha-\alpha^2}.
\label{eq:theta}
\eeq

Thus, Eqs. (\ref{eq:Jr0}) and (\ref{eq:rhor0}) together with the solution of $\rho_0$ from 
\ref{eq:theta} provides the complete mean field solution   of the  asymmetric exclusion process 
on the Cayley tree.  
To check the validity of the MF theory, we  
simulate this exclusion process   with $\alpha=0.7$ and $\beta=0.2$ on a  Cayley tree ($q=2$)with 
$N=6,8,10,12$ and plot $\rho_i$ versus $i$ in semi-log  scale ( Fig. \ref{fig:rhoN}).  
Clearly the densities $\rho_i$, except  $\rho_{N-1}$, show an exponential decay which 
agree quite well with the MF results (\ref{eq:rhor0}). The plot depicts that the  finite 
size corrections are absent here;  $i.e.,$  the level density of, say at level $3$ 
($i.e. \rho_3$), is independent of the system size $N=6,8,10,12$.   Further, it  appears
that $\rho_{N-1}$ varies with system size  $N$ as $\rho_{N-1} =  c/2^{N-1}$,  with 
$c>\rho_0.$ 

To obtain the boundary density $\rho_{N-1}$, we use the second boundary condition (\ref{eq:eqN}); 
in  steady state $J_{N-2}  = \beta\rho_{N-1}$. Since $J_{N-2} = I_q/ q^{N-1}$ [from Eq. \ref{eq:Ji}], 
 we have 
 \beq \rho_{N-1}=\frac{c}{q^{N-1}} ~~{\rm with}~~ c = \frac {I_q}{\beta} =\frac{\alpha(1-\phi_0)}{\beta}.\label{eq:rhoN-1}
\eeq
In the inset of Fig. \ref{fig:rhoN}, $c$ versus $\beta$ obtained from numerical simulations 
for  two different values of $\alpha=0.4, 0.7$  are  compared with corresponding mean field  values 
given by Eq. (\ref{eq:rhoN-1}).

\begin{figure}
 \centering
 \includegraphics[height=6cm]{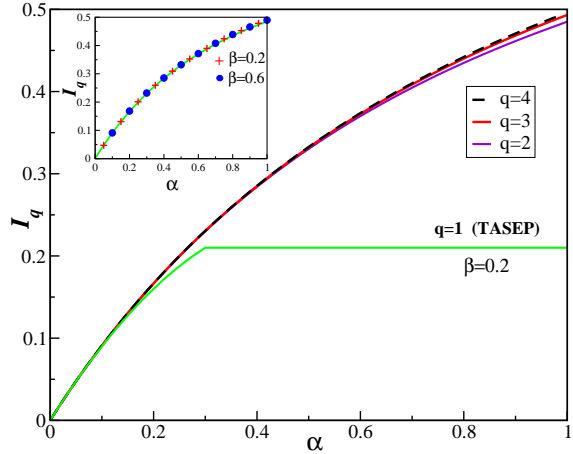}
 \caption{  Plot of $J_q$ verses $\alpha$ with different $q=1, 2,3,4.$. 
For $q=1$ (TASEP) the current $J_1$ depends on $\beta$ which is taken to be $\beta=0.2$ 
Inset : For $q=2$, $J$ versus $\alpha$  obtained from numerical simulations with $\beta=0.2,0.6$ 
(symbols) is compared with Eq. \ref{eq:Jr0}. }
 \label{fig:jq}
\end{figure}

For a generic Cayley tree with $q\ge 2$, the current  $I_q = \alpha (1-\rho_0)$  where 
$\rho_0$ is given by Eq. (\ref{eq:theta}). Evidently,   $I_q$ is independent of the exit 
rate $\beta$ and increases monotonically with $\alpha$. To verify this we calculate 
$I_q$ for a Cayley tree with $q=2$ using Monte-Carlo simulations  for two different values of 
$\beta= 0.2, 0.6 $ and plot them against  $\alpha$ (inset of Fig. \ref{fig:jq}). 
The mean field current (\ref{eq:Jr0}) for $q=2$ , draws as a solid line  there, 
shows an excellent agreement.  Current $I_q$ for $q>2$ are shown in the main 
figure [results  from the simulations data (not shown here) matches very well with  
Eq.(\ref{eq:Jr0})].

Figure \ref{fig:jq} compares $I_q$  for  different $q=1,2,3,4$.  It is only 
$I_1$, the current in TASEP, that depends on  $\beta$ (chosen here as $\beta=0.2$). 
As expected, $I_q$  is a strictly non-decreasing function 
of the entry rate $\alpha$.  Thus, the maximum achievable current 
on a Cayley tree is  $I_q(\alpha=1)$. For example, when $q=2$,  
the maximum current 
\beq
I^{max}_2={1\over3}\left[1+10cos\right( \frac \pi 3  
+ \frac 1 3 tan^{-1} (\frac{6\sqrt{426}}{17} ) \left)\right]= 0.485 
\eeq
is almost twice as large as that of TASEP ($q=1$)\cite{note1}. 
From Fig.  \ref{fig:jq}, it appears that $\lim_{q \to \infty} I_q(\alpha=1)  = {1\over 2}$. 
This can be understood  from the fact that  when   $q \to \infty$, the
rate of out flow from  the root $j=0$  is unity,  as one of the the 
infinitely many neighbours  of the root are expected to be vacant with probability $1$.
Thus the particle density $\rho_0$ at the root  is expected to be $1/2$  for $\alpha=1$, 
resulting in $I_\infty=1/2$. Another interesting fact about $I_q$ , is the  
following in-equality, 
\beq
I_1(\alpha,\beta) < I_2(\alpha)\dots <I_\infty(\alpha).\nonumber
\eeq
 which holds for any  arbitrary value of $(\alpha, \beta)$.

In conclusion, we have studied  the asymmetric exclusion process on a regular  Cayley tree 
with arbitrary  coordination number $z=q+1$, where particles  enter the system only at 
the parent site with rate $\alpha$, and  exit from one of the sites at the last level 
with rate $\beta$. In the bulk they move from  occupied sites to one of their downward 
unoccupied neighbours chosen randomly. TASEP is a  special case of this model for $q=1$ which exhibit  
boundary driven phase transitions. In this case the steady state current $I_q$ 
is different in different phases and depend both on $\alpha$ and  $\beta$. 
Surprisingly, 
for $q\ge2$, we find that there is only  one phase and current $I_q$ is 
independent of $\beta$. Further, $I_q$ increases monotonically  with $\alpha$ and  
reaches a maximum value twice as large as that of TASEP  even for small 
coordination numbers (say, $q=2$). 
Again, the density profiles  do not show any boundary layers or  finite size 
corrections.  It is like a free flow of particles from one end to the  other; 
only the last level is effected by the exit rate.
The model could find application in nutrient transport in 
blood-circulation systems, air circulation in lung or in antibody  transport
in immune systems or water transport in trees. Possibly, branching is essential 
in all these systems (say trees) as the  flow (of water at different levels) need 
to be maintained independent of the exit rates (weather conditions) at the last 
level(leaves).

\end{document}